\documentclass[aps,pra,superscriptaddress]{revtex4-2}
\usepackage[utf8]{inputenc}
\usepackage{amsmath, amsfonts}

\usepackage{amscd}
\usepackage{amsthm}
\usepackage{amssymb}
\usepackage{float}
\usepackage{cases}
\usepackage{mathtools}
\usepackage{graphicx}
\usepackage{physics}

\usepackage{xcolor}
\definecolor{algogreen}{RGB}{46, 106, 107}
\usepackage[colorlinks=true,
bookmarks=false,
linkcolor=algogreen,
urlcolor=algogreen,
citecolor=algogreen,
breaklinks]{hyperref}

\usepackage{natbib}

\usepackage[normalem]{ulem}
\usepackage{xcolor}

\begin{document}

\title{Link prediction with continuous-time classical and quantum walks}
\author{Mark Goldsmith}
\author{Guillermo García-Pérez}
\author{Joonas Malmi}
\author{Matteo A. C. Rossi}
\author{Harto Saarinen}
\author{Sabrina Maniscalco}
\affiliation{Algorithmiq Ltd, Kanavakatu 3 C, FI-00160 Helsinki, Finland}

\date{\today}

\begin{abstract}
Protein-protein interaction (PPI) networks consist of the physical and/or functional interactions between
the proteins of an organism. Since the biophysical and
high-throughput methods used to form PPI networks are expensive,
time-consuming, and often contain inaccuracies, the resulting networks
are usually incomplete. In order to infer missing interactions in these networks,
we propose a novel class of link prediction methods based on
continuous-time classical and quantum random walks. In the case of
quantum walks, we examine the usage of both the network adjacency and
Laplacian matrices for controlling the walk dynamics. 
We define a score function based on the corresponding transition probabilities and perform tests on
four real-world PPI datasets.
Our results show that continuous-time classical random walks and
quantum walks using the network adjacency matrix can successfully
predict missing protein-protein interactions, with performance
rivalling the state of the art.
\end{abstract}

\maketitle

\section{Introduction}
The link prediction problem has long been an active area of research, with applications ranging from friendship recommendation in social networks~\cite{adamic2003friends, murata2007link, leskovec2010predicting} to finding missing interactions between proteins~\cite{kovacs2019network, Liu2010LocalRW}. In this paper, we are interested in the latter. 

One particularly successful class of link prediction methods are those based on random walks \cite{Liu2010LocalRW, Che2021RWR, Zhou2021biasedRWR}.
Random walk algorithms have been explored more generally throughout the field of network science, and many different applications exist. These include the ranking of webpages using PageRank~\cite{pagerank, das2013fast}, collaborative filtering~\cite{fouss2007random}, and computer vision~\cite{pan2004automatic}. Many random walk link prediction algorithms have also been studied~\cite{Liu2010LocalRW, tong2006fast}. These methods typically rely on discrete-time random walks. 

In contrast, in this paper we propose a class of link prediction methods based on continuous-time random walks. 
Moreover, the continuous-time setting allows us to propose a new link prediction method using quantum walks, which closely resembles the classical method described here.

Continuous-time quantum walks, initially proposed in~\cite{Farhi1998}, are the quantum analogues of continuous-time classical random walks, which describe the propagation of a particle over a discrete set of positions. Together with their discrete-time counterpart~\cite{Aharonov1993}, they have received a lot of attention for their applications in quantum information processing~\cite{kempe2003quantum, venegas2012quantum}, quantum computation~\cite{childs2009universal}, and quantum transport~\cite{mulken2011continuous}.
However, only a few recent methods have attempted to use quantum walks for link prediction, using their discrete-time~\cite{Qian2017LPbyDTQW}, and continuous-time \cite{Omar2021} variations. 


In order to evaluate our proposed methods, we have conducted experiments on several networks, and have found that both the classical and quantum walks outlined here are particularly good at finding missing links in protein-protein interaction (PPI) networks.
Protein-protein interactions play a critical role in all cellular processes, ranging from cellular division to apoptosis. Elucidating and analyzing PPIs is thus essential to understand the underlying mechanisms in biology and, eventually, to unveil the molecular roots of human disease \cite{PPI3}. Indeed, this has been a major focus of research in recent years, providing a wealth of experimental data about protein associations \cite{PPI1,rolland2014proteome}. Current PPI networks, called interactomes, have been constructed using a number of techniques, but despite the enormous advancement, the current coverage of PPIs is still rather poor (for example, it is estimated that only around 10\% of interactions in humans are currently known~\cite{luck2020reference}). Additionally, despite considerable improvements in high-throughput (HTP) techniques, they are still prone to spurious errors and systematic biases, yielding a significant number of false-positives and false-negatives. This limitation impedes our ability to assess the true quality and coverage of the interactome.



Recently, a number of algorithms have been developed to predict protein-protein interactions. 
In a recent study by Kovács et al.~\cite{kovacs2019network}, a novel PPI-specific link predictor was proposed. Their link predictor is biologically motivated by the so-called L3 principle, and it was shown to be superior to other general link predictors when applied to PPI data. The exceptional success of the L3 framework is rooted in its ability to capture the structural and evolutionary principles that drive PPIs. The results of Kovács and collaborators prove that,  contrary to the current network paradigm, interacting proteins are not necessarily similar and similar proteins do not necessarily interact, questioning the traditional validation strategy based on biological similarity of the predicted protein pairs.

However, the L3 link prediction method, considered the most successful to date for PPIs, as well as most other existing link prediction methods, are not without limitations. The most common approaches cannot find interacting partners for proteins without known links, self-interacting proteins, or links between proteins that have longer paths between them. Given the low coverage of the current PPI databases, this can be a significant drawback. It is therefore highly desirable to complement the existing frameworks with methods relying on the exploration of the whole network, and consequently be able to incorporate information provided by longer paths. This is one of the the main goals of our paper as we propose two novel random-walk based link prediction methods.


\section{The method}
Consider a network modelled by an undirected graph $G=(V,E)$, where $V$ is the set of nodes of size $n$ and $E$ is the set of edges. We allow for the existence of self-edges, so that for any node $i$, the edge $(i,i)$ may or may not be present in $E$. The \emph{adjacency matrix} of $G$ is the $n \times n$ matrix defined by 
\begin{equation*}
    A = (A_{ij}) =
    \begin{cases}
    1, & \text{ if } (i,j) \in E, \\
    0, & \text{ if } (i,j) \not \in E.
    \end{cases}
\end{equation*}
The \emph{graph Laplacian} is defined as $L=D-A$, where $D$ is the \emph{degree matrix} defined by $D = \text{diag} \left(\sum_{j}A_{1j}, \ldots, \sum_{j}A_{nj}\right)$.

The precise details of the random walks we employ will be described in the next subsections. For now, it suffices to consider the notion of a probability transition matrix that evolves over time, denoted by $P(t)$; for a graph $G$, the probability of the random walker being at node $v$ at time $t$, given that it began at node $u$, is thus $P_{uv}(t)$.
For a fixed time $t$, we define the score $S(i,j;t)$ between two non-adjacent nodes $i$ and $j$ at time $t$ to be

\begin{numcases}{S(i,j;t) = } P_{ij}(t)\left(k_i + k_j\right) \quad i \ne j \label{notself} \\
\frac{1}{2}\sum_{u \in N(i)} P_{iu}(t) \quad \, \, i = j, \label{self}
\end{numcases}
where $N(v)$ denotes the neighbourhood of node $v$, and $k_v = \sum_j A_{vj}$ is the degree of node $v.$ Equations~(\ref{notself}) and~(\ref{self}) handle the cases of distinct nodes and self-edges, respectively.
While the score in Equation~(\ref{notself}) is superficially similar to the one proposed in~\cite{Liu2010LocalRW}, the fact that we are using continuous-time random walks leads to several key differences: the continuous-time nature of our method allows for a wider range of time parameters $t$ to use; in the continuous-time setting there is symmetry in the transition probabilities, i.e. $P_{ij}(t) = P_{ji}(t)$ for all nodes $i,j;$ and finally, there is a close relationship in the implementation of classical and quantum walks in the continuous-time setting.

Regardless of which type of random walk is used, we must choose a value $t$, representing the time-duration of the walk. We start the walk at time $t_0 = 0$, and let it run for a time $t$, at which point we extract the scores for the target edges from the probability distributions. In the case of a continuous-time classical random walk, the expected time it takes for a random walker to leave a node $i$ is $1/k_i.$ This motivates the idea that the amount of time we let the random walk run should be related to the degree distribution of the network. Our experiments show that setting $t = 1/\langle k \rangle$, where $\langle k \rangle$ is the average node degree in the graph, yields good results for both classical and quantum walks.

Next, we describe the different type of walks we use in more detail.

\subsection{Continuous-time random walks}
A continuous-time (classical) random walk (CRW) is a Markov process with state space $V$ characterized by a rate matrix $Q$ and initial distribution $\mathbf{p}(0)$ over the set of nodes. Here, we consider edge-based random walks~\cite{masuda2017random} (as opposed to node-based), which are characterized by setting $Q=-L$, where $L$ is the Laplacian of the underlying graph. In this case, the dynamics are governed by the equation 
\begin{equation}\label{eq:RWprobabilitymatrix}
    \mathbf{p}(t) = \mathbf{p}(0)e^{-tL}.
\end{equation}

Intuitively, the random walker operates as follows. Every edge of the graph is associated with an independent Poisson process with unit intensity. When the walker is at some node, it will remain there until one of the Poisson processes at an incident edge jumps, at which point the walker follows that edge to the corresponding neighbour, and the process repeats. Note that this implies that, on average, a random walker will spend less time waiting at a higher degree node than at a lower degree node.




\subsection{Continuous-time quantum walks}

In contrast to a classical random walk, a quantum walk on a network evolves according to the laws of quantum physics. A major implication of this is that the trajectories of the walker across the network can interfere constructively or destructively. This interference causes the evolution of the quantum walker to sometimes be significantly different from the classical one \cite{Aharonov1993, Difference-between-QW-and-CW}.

A continuous-time quantum walk (QRW) \cite{Farhi1998} on a graph $G$ is defined by considering the Hilbert space $\mathcal{H}$ spanned by the orthonormal vectors $\{\ket{i}\}_{i=1}^n$, corresponding to the $n$ nodes of the graph and the unitary transformation $e^{-itH}.$ This transformation implies that the state vector in $\mathcal{H}$ at a time $t$ after starting from initial time $t_0 = 0$ is given by the evolution 
\begin{equation}\label{eq:QWstatevector}
\left| \psi (t)\right\rangle = U (t) \left| \psi (0)\right\rangle,
\end{equation}
where $U(t) = e^{-itH}$ is the unitary evolution operator, and $H$ is the Hamiltonian. In general, the Hamiltonian $H$ can be almost any Hermitian matrix related to $G$ as long as it describes the structure of the network \cite{venegas2012quantum}, but the most common choices are the Laplacian $L$ or the graph adjacency matrix $A$ \cite{LvsAQuantumWalk}. In this paper we will use both the Laplacian and the adjacency matrix as the Hamiltonian separately, and therefore can compare different realizations of quantum walks for the link prediction task.

In order to obtain a probability transition matrix analogous to the one in Eq.~(\ref{eq:RWprobabilitymatrix}), we must take the square of the modulus of the entries of $U(t)$. The entries of the probability transition matrix are given by
\begin{equation}
\label{eq:QRWprobabilitymatrix}
    P_{ij} (t)= |\langle j | e^{-itH} | i \rangle|^2.
\end{equation}
These transition probabilities can then be used to compute scores for missing edges as described in Equations~(\ref{notself}) and~(\ref{self}) above.

We emphasize that the usage of continuous-time quantum walks for link prediction is a new direction of research, with very few studies conducted so far. The method proposed in \cite{Omar2021}, in particular, appears to be competitive with some state-of-the-art link prediction methods in certain real networks. While some aspects of their algorithm are similar to the quantum version of our random walk algorithm, the implementation details and calculation of the link prediction scores are very different. Moreover, their algorithm requires entanglement with an additional ancilla. While this would be feasible in a hypothetical implementation on a quantum computer, the typical sizes of relevant real networks are way beyond the capabilities of current and near-term quantum hardware. Simulations on classical computers are required, but the presence of the extra ancilla increases the complexity of simulations.

\section{Results}
In this section we first describe the datasets and networks that were used for evaluation, then we compare our methods to some well-known and state-of-the-art link prediction algorithms. 

\subsection{Datasets and evaluation}
\label{data}

\begin{figure}[H]
    \centering
    \includegraphics[width=0.75\linewidth]
    {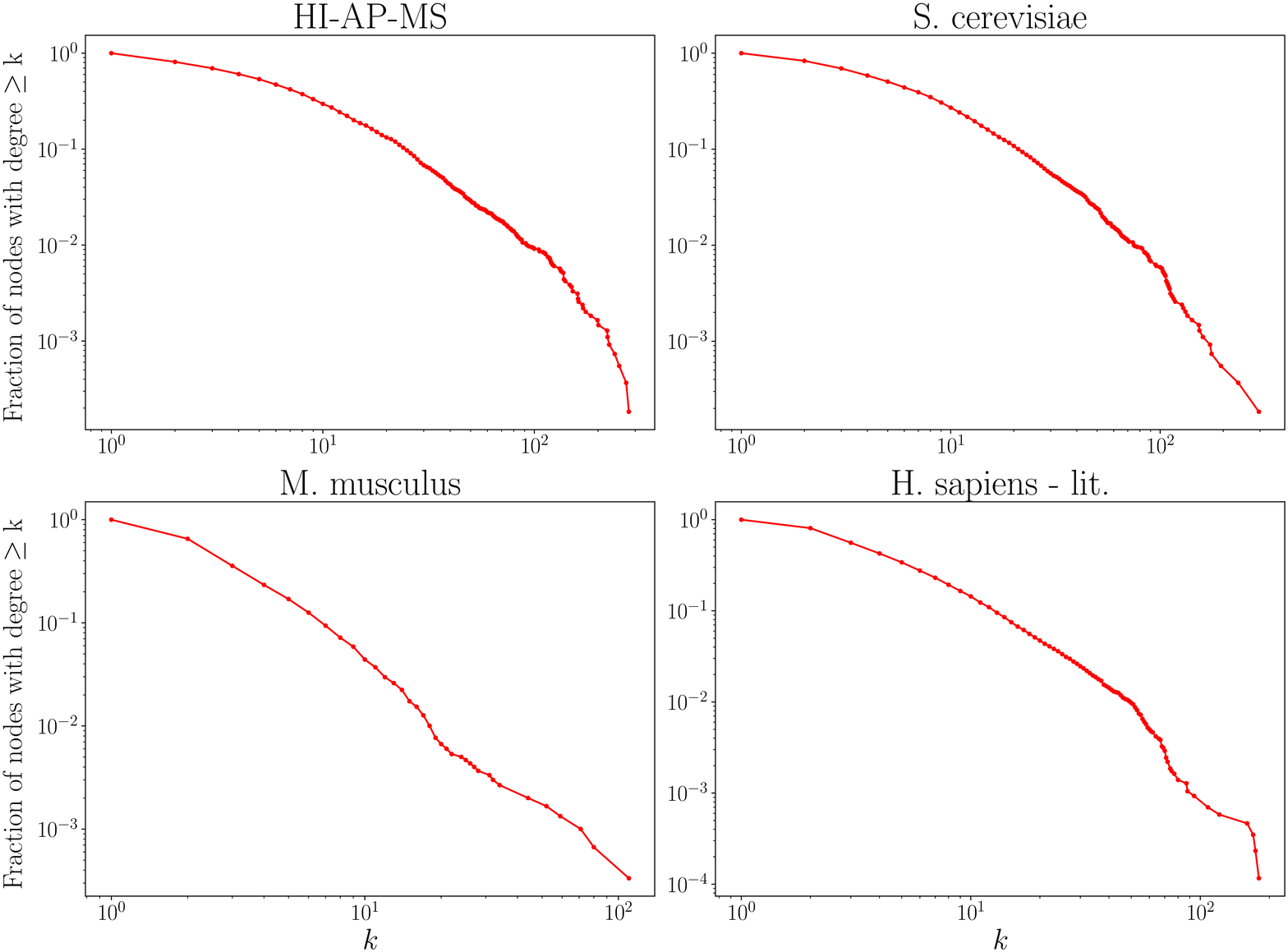}
    \caption{\textbf{Complementary cumulative degree distributions.} For each degree value $k$ ($x$-axis), the proportion of nodes with degree greater than or equal to $k$ ($y$-axis) is shown, each on a logarithmic scale.}
    \label{fig:degdist}
\end{figure}

We tested our link prediction methods on four different PPI networks. Three of them are high-quality PPI networks from the HINT database \cite{pombe}: M. musculus and S. cerevisiae are networks resulting from combining data from high-throughput screenings with literature-curated small-scale experiments, and H. sapiens consists of only the latter, representing a subset of the human interactome. The fourth network we tested on, HI-AP-MS, is another large-scale map of the human interactome, but in this case was produced by affinity purification followed by mass spectrometry~\cite{hein2015human}. Some statistics of these networks are listed below in Table~\ref{tab:network_properties}, and their degree distributions are shown in Figure \ref{fig:degdist}.


\begin{table}[ht]
\centering
\begin{tabular}{|l|ccccccc|}
\hline
Network &   $|V|$ &    $|E|$ & $\langle k \rangle$ & $\rho$ &  $C$ &     $A$ &  \text{SIPs} \\
\hline
HI-AP-MS          &  5457 &  28780 &   10.548 &            0.002 &        0.158 &         -0.188 &        1127 \\
S. cerevisiae     &  5420 &  25035 &    9.238 &            0.002 &        0.122 &         -0.121 &        1417 \\
M. musculus       &  2995 &   4671 &    3.119 &            0.001 &        0.103 &         -0.070 &         978 \\
H. sapiens &  8601 &  24627 &    5.727 &            0.001 &        0.171 &          0.138 &        4409 \\
\hline
\end{tabular}
\caption{\textbf{Some properties of the networks that were tested}. $|V|:$ number of nodes, $|E|:$ number of edges, $\langle k \rangle :$ average degree, $\rho:$ network density, $C:$ average clustering, $A:$ assortativity, $\text{SIPs}$: number of self-interacting proteins (self-edges).}
\label{tab:network_properties}
\end{table}

For each dataset, we randomly removed $P\%$ of the edges in the original
network, for $P \in \{10, 20, 30, 40, 50\}$ and reserved these edges as positive test cases. All of the nonexistent edges were used as negative testing data. These
positive and negative edges were used for evaluation, and the
remaining $(100-P)\%$ existing edges were used for running the models in
question. This process was repeated 20 times for each $P$, and the
results were averaged (see results in the next subsection).

\subsection{Comparison to other methods}
In order to test our methods, we selected 5 other popular link prediction methods to compare against: 
\textit{L3} relies on a weighted counting of paths of length three, and was designed specifically to predict links in PPI networks~\cite{kovacs2019network};
\textit{preferential attachment} (PA) defines a score between two disconnected nodes by multiplying their degrees~\cite{liben2007link, barabasi2002evolution};
\textit{common neighbours} (CN) is a straightforward heuristic that assigns a score to edge $(u,v)$ defined by the number of neighbours that $u$ and $v$ have in common; 
\textit{Adamic-Adar} (AA) is an adaptation of the common neighbours idea, but adds more weight to less connected neighbours \cite{adamic2003friends};
\textit{SPM} is based on perturbations of the adjacency matrix of the graph \cite{Lu2015}.

The following tables show average precision \cite{precision} and area under the receiver operating characteristic (ROC) curve \cite{auc} values for the 4 different networks described in Section~\ref{data}. Each value is averaged over 20 runs (20 randomly selected edges removals). We compare three variations of our proposed methods, labelled as `QRW-A', `QRW-L', and `CRW', referring to quantum random walks using the network adjacency matrix as Hamiltonian, quantum random walks using the network Laplacian matrix as Hamiltonian, and classical random walks, respectively.

\begin{table}[!ht]
\centering
\begin{tabular}{|r||lll|lllll|}
\hline
Network &  QRW-A &  QRW-L &             CRW &     L3 &     PA &     CN &     AA &    SPM \\ \hline
S. cerevisiae     &  0.877 &  0.875 &  \textbf{0.881} &  0.874 &  0.828 &   0.78 &  0.781 &            0.84 \\
HI-AP-MS          &  0.875 &  0.879 &  \textbf{0.885} &  0.879 &  0.866 &  0.737 &  0.737 &           0.855 \\
H. sapiens - lit. &  0.852 &  0.853 &           0.854 &   0.86 &  0.732 &  0.804 &  0.803 &  \textbf{0.862} \\
M. musculus       &   0.73 &   0.73 &           0.731 &   0.73 &  0.563 &  0.683 &  0.677 &  \textbf{0.764} \\
\hline
\end{tabular}
\caption{Area under the ROC curve for 10\% edge removals, averaged over 20 trials.}
\end{table}

\begin{table}[!ht]
    \centering
   \begin{tabular}{|r||lll|lllll|}
    \hline
    Network &  QRW-A &  QRW-L &             CRW &     L3 &     PA &     CN &     AA &    SPM \\ \hline
    S. cerevisiae     &  0.839 &  0.838 &  \textbf{0.843} &  0.799 &   0.81 &  0.689 &  0.689 &  0.778 \\
    HI-AP-MS          &  0.837 &   0.84 &  \textbf{0.846} &  0.792 &  0.843 &  0.664 &  0.665 &  0.771 \\
    H. sapiens - lit. &  0.802 &  0.802 &  \textbf{0.803} &   0.76 &  0.733 &  0.708 &  0.705 &  0.792 \\
    M. musculus       &  0.686 &  0.686 &  \textbf{0.687} &   0.63 &    0.6 &  0.611 &  0.603 &  0.673 \\
    \hline
    \end{tabular}
    \caption{Area under the ROC curve for 50\% edge removals, averaged over 20 trials.}
    \label{auc50}
\end{table}

\begin{table}[!ht]
\centering
\begin{tabular}{|r||lll|lllll|}
\hline
Network &  QRW-A &  QRW-L &             CRW &     L3 &     PA &     CN &     AA &    SPM \\ \hline
S. cerevisiae     &           0.035 &  0.033 &  \textbf{0.053} &  0.038 &  0.003 &  0.022 &  0.028 &           0.047 \\
HI-AP-MS          &           0.078 &  0.035 &            0.08 &  0.078 &  0.003 &  0.034 &  0.046 &  \textbf{0.123} \\
H. sapiens - lit. &  \textbf{0.121} &  0.115 &           0.111 &   0.08 &  0.003 &  0.074 &  0.089 &             0.1 \\
M. musculus       &           0.034 &  0.029 &           0.032 &  0.027 &  0.001 &  0.012 &  0.015 &  \textbf{0.047} \\
\hline
\end{tabular}
\caption{Average precisions for 10\% edge removals, averaged over 20 trials.}
\label{ap1}
\end{table}

\begin{table}[!ht]
\centering
\begin{tabular}{|r||lll|lllll|}
\hline
Network &  QRW-A &  QRW-L &             CRW &     L3 &     PA &     CN &     AA &    SPM \\
\hline
S. cerevisiae     &    0.1 &  0.085 &           0.111 &  \textbf{0.113} &  0.013 &  0.054 &  0.068 &           0.085 \\
HI-AP-MS          &  0.159 &  0.074 &           0.154 &  \textbf{0.165} &  0.015 &  0.065 &   0.08 &           0.162 \\
H. sapiens - lit. &  0.185 &  0.181 &           0.194 &           0.169 &  0.011 &  0.136 &  0.157 &  \textbf{0.217} \\
M. musculus       &  0.065 &  0.056 &  \textbf{0.067} &           0.058 &  0.003 &  0.025 &  0.031 &           0.017 
\\
\hline
\end{tabular}
\caption{Average precisions for 50\% edge removals, averaged over 20 trials.}
\end{table}

\begin{figure}[!ht]
    \centering
    \includegraphics[width=0.81\linewidth]{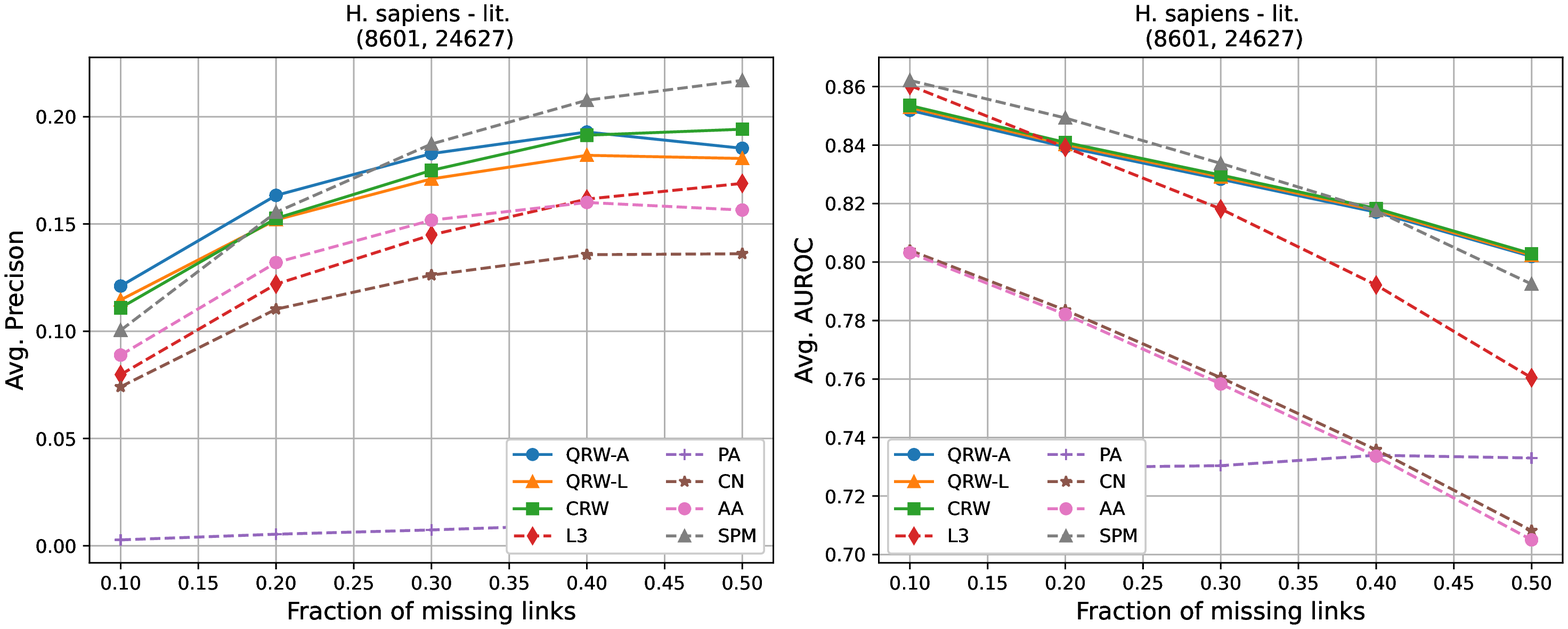}
    \caption{Average precisions (left) and area under the ROC curve (right) as a function of the fraction of true links that have been removed from the literature-curated Homo sapiens network found in~\cite{pombe}. Plotted values are the averages over 20 trials.}
    \label{fig:HsapAUC}
\end{figure}

\begin{figure}[!ht]
    \centering
    \includegraphics[width=0.81\linewidth]{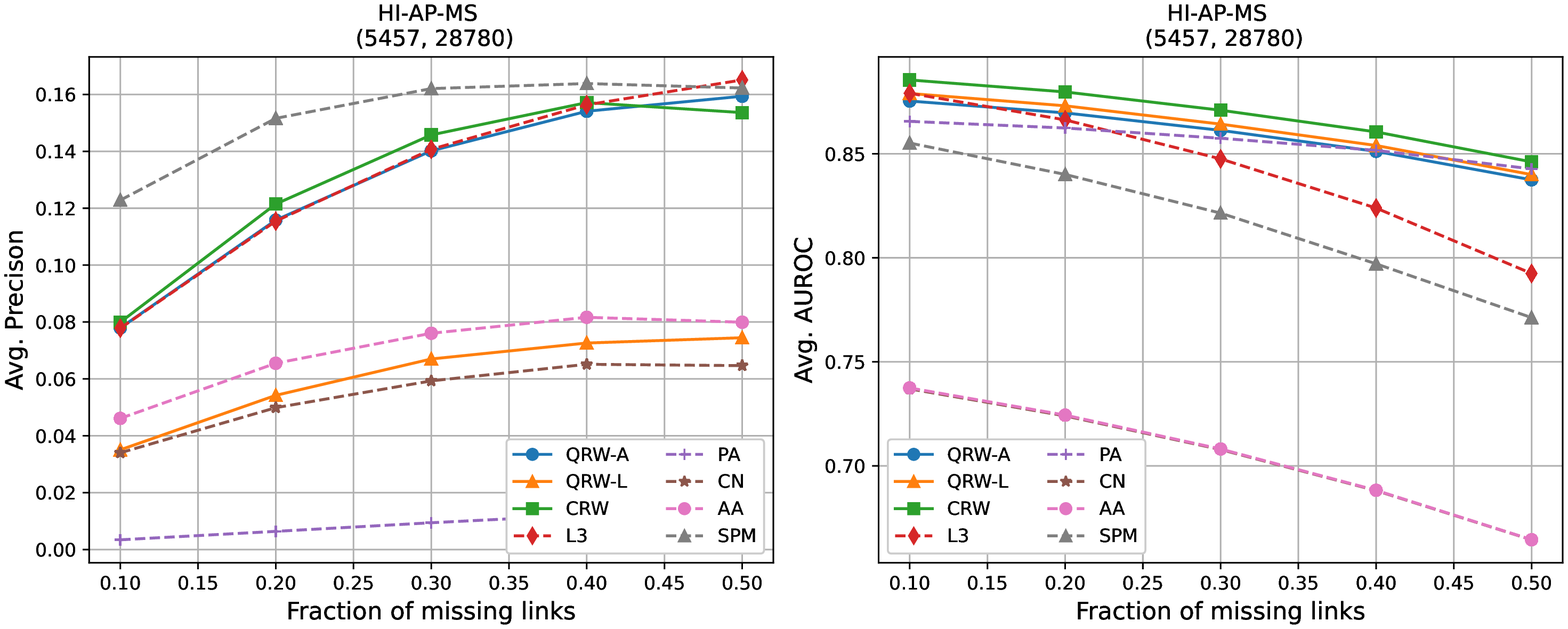}
    \caption{Average precisions (left) and area under the ROC curve (right) as a function of the fraction of true links that have been removed from the Homo sapiens network found in~\cite{hein2015human}. Plotted values are the averages over 20 trials.}
    \label{fig:hiapmsAUC}
\end{figure}

\begin{figure}[!ht]
    \centering
    \includegraphics[width=0.81\linewidth]{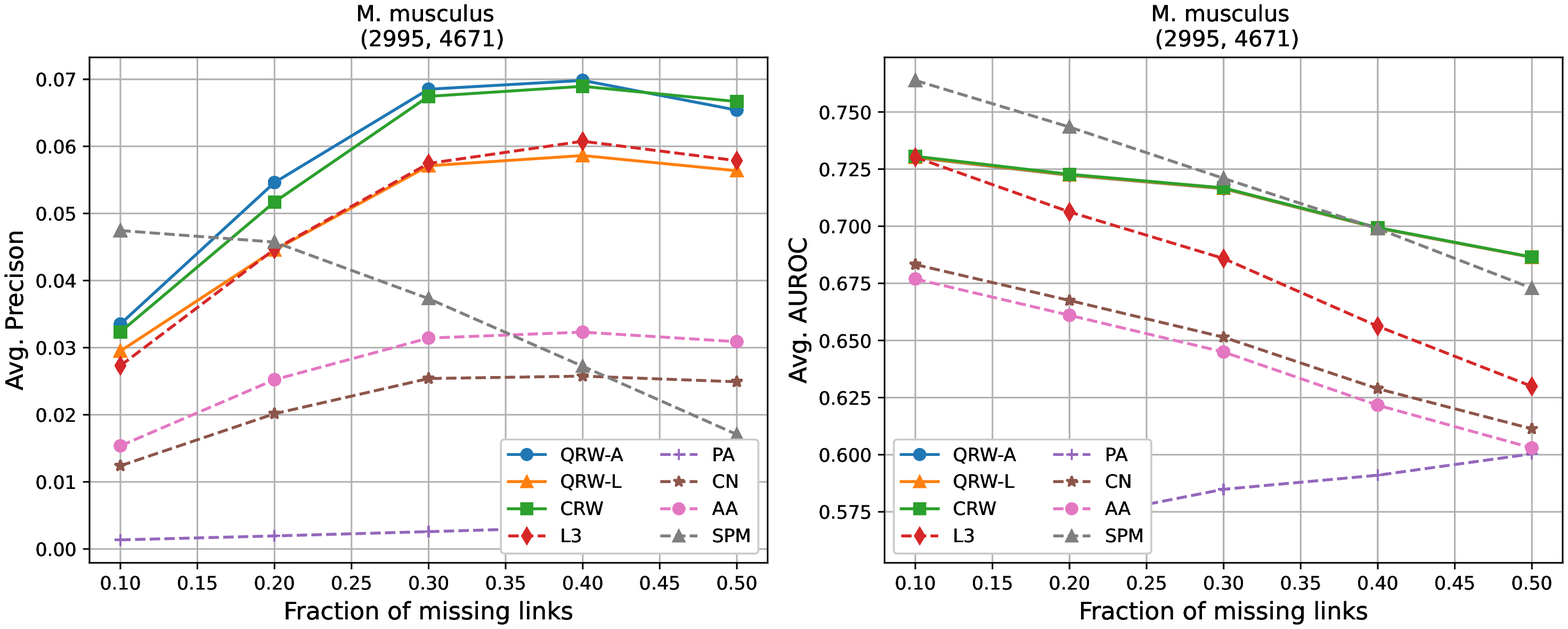}
    \caption{Average precisions (left) and area under the ROC curve (right) as a function of the fraction of true links that have been removed from the Mus musculus network found in~\cite{pombe}. Plotted values are the averages over 20 trials.}
    \label{fig:musculusAUC}
\end{figure}

\begin{figure}[!ht]
    \centering
    \includegraphics[width=0.81\linewidth]{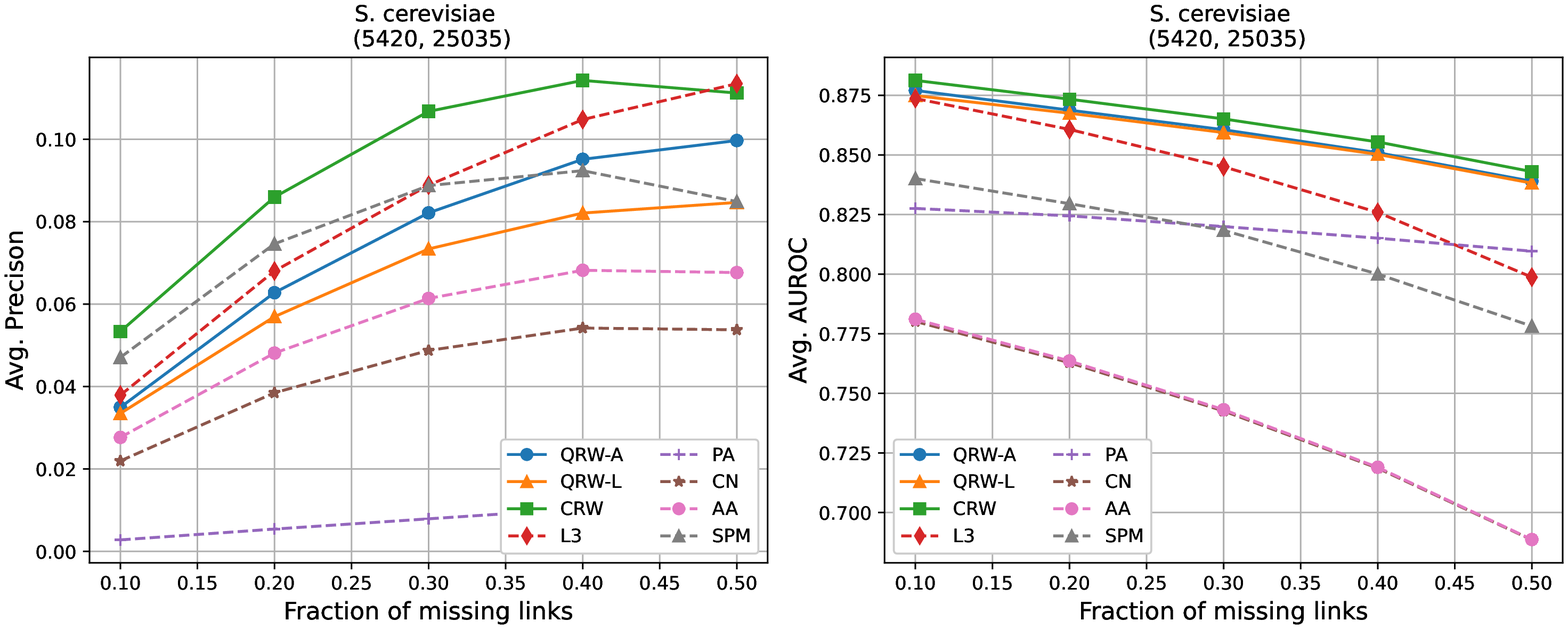}
    \caption{Average precisions (left) and area under the ROC curve (right) as a function of the fraction of true links that have been removed from the Saccharomyces cerevisiae network found in~\cite{pombe}. Plotted values are the averages over 20 trials.}
    \label{fig:cereviAUC}
\end{figure}

For completeness, in Figures \ref{fig:HsapAUC}-\ref{fig:cereviAUC} we also include plots showing the relationship of average precision and area under the ROC curve as a function of edge removal fraction.

\section{Discussion}
The experimental results in the previous section show that our methods perform well on a variety of PPI networks. In particular, we see that our classical random walk method yields the best performance of all algorithms tested with respect to the area under the receiver operating characteristic curve when 50\% of the edges are removed (Table~\ref{auc50}). The quantum walks using the adjacency Hamiltonian yield the best average precision on one of the human proteome datasets when 10\% of edges are removed (Table~\ref{ap1}). Furthermore, the adjacency Hamiltonian almost always beats out the Laplacian as the better choice when comparing the results of quantum walks. However, as is usually the case with link prediction methods, there is no single method that wins in all cases. One must pay careful attention to the goals of the application in question and decide on which metric to rely on accordingly.

Finally, we mention a few points about the computational complexity of our algorithm and its implementation. The bottleneck of our algorithm, in either the classical or quantum case, is the computation of the matrix exponential appearing in Equation~(\ref{eq:RWprobabilitymatrix}) and Equation~(\ref{eq:QRWprobabilitymatrix}), which is a very well-studied problem with a long history~\cite{moler2003nineteen}. Our experiments were done using the `matrix\_exp' function in PyTorch~\cite{NEURIPS2019_9015},
which is an implementation of the Taylor polynomial approximation algorithm described in~\cite{bader2019computing}. The problem is thus reduced to a constant number of matrix multiplications, another well-studied problem that can be solved more quickly than the naive $O(n^3)$ method; for example, in $O(n^{2.376})$ time using the Coppersmith–Winograd algorithm~\cite{coppersmith1982asymptotic}. It is also worth noting that in this implementation of matrix exponentiation, and many others, the norm of the matrix being exponentiated has an impact on running time, so that using a small $t$, as tends to be the case in our algorithm, may help in this regard.

\section{Conclusions}
Although experimental methods have greatly improved in the past ten years, most interactomes remain far from being complete. It is therefore important to discover new computational methods for inferring interactions from incomplete datasets. We have described a class of algorithms based on continuous-time random walks that rank among the best link prediction methods tested on PPI networks.

Furthermore, the quantum continuous-time random walks described here are among the first successful quantum-inspired link prediction methods.
Although we have found that using the reciprocal of the average degree provides a good time length for which to run the random walks, many further options can still be explored: using cross-validation to choose a more optimal value, or using times that depend on the walker's location are immediate candidates.
Another open direction of research involves the choice of Hamiltonian. Our experimental results demonstrate a strong sensitivity on the Hamiltonian used for controlling the quantum walks. While the adjacency matrix yields better results than the Laplacian on most of the networks we tested, it would be beneficial to understand why this is the case. This also indicates the potential for improvement if better Hamiltonians can be found for the purpose of link prediction. Further investigations in this direction may yield better methods and insights into both the networks being studied, and the quantum walks being employed.

\textbf{Competing interests:} The authors declare no competing interests.
\textbf{Authors contributions:} MG, HS conceived the algorithm.
GGP, MACR, SM designed and directed the research.
MG, JM implemented the algorithms and ran simulations.
MG, JM, HS wrote the first version of the manuscript.
All authors contributed to scientific discussions and to the writing of the manuscript.

\bibliographystyle{acm}
\bibliography{bibliography}

\end{document}